\documentclass{article}
\usepackage{ijcai20}

\usepackage{times}
\usepackage{soul}
\usepackage{url}
\usepackage[hidelinks]{hyperref}
\usepackage[utf8]{inputenc}
\usepackage[small]{caption}
\usepackage{graphicx}
\usepackage{amsmath}
\usepackage{amsthm}
\usepackage{booktabs}
\usepackage{float}
\usepackage{multirow}
\usepackage{diagbox}
\usepackage[linesnumbered,titlenumbered,ruled,vlined,commentsnumbered,noend]{algorithm2e}

\usepackage{xcolor}
\usepackage{comment}
\usepackage{color}

\usepackage{epsfig}
\usepackage{graphicx}
\usepackage{amsmath}
\usepackage{amssymb}

\usepackage{multirow}
\usepackage{enumitem}
\usepackage{moresize}
\usepackage{epstopdf}
\usepackage{array}

\usepackage{makecell}

\usepackage{wrapfig}
\usepackage{subfigure}
\usepackage[hang,flushmargin]{footmisc} % remove footnote indentation

\usepackage[font=footnotesize]{caption}

\usepackage[belowskip=-10pt,aboveskip=5pt]{caption} % re-arrange tables on last page, select

\usepackage{xspace}

\usepackage{xspace}
\makeatletter
\DeclareRobustCommand\onedot{\futurelet\@let@token\@onedot}
\def\@onedot{\ifx\@let@token.\else.\null\fi\xspace}
\def\eg{\emph{e.g}\onedot} 
\def\ie{\emph{i.e}\onedot} 
 
\def\etc{\emph{etc}\onedot} 
 
\def\etal{\emph{et al}\onedot}
\makeatother

\definecolor{darkorange}{rgb}{1.0, 0.55, 0.0}
\definecolor{lincolngreen}{rgb}{0.11, 0.35, 0.02}
\definecolor{cornflowerblue}{rgb}{0.39, 0.58, 0.93}
\definecolor{cobalt}{rgb}{0.0, 0.28, 0.67}

\title{\emph{FakeSpotter}: A Simple yet Robust Baseline for Spotting AI-Synthesized Fake Faces}

\author{
Run Wang$^1$\footnote{Corresponding author, E-mail: \href{mailto:runwang1991@gmail.com}{runwang1991@gmail.com}}\and
Felix Juefei-Xu$^2$\and 
Lei Ma$^3$\and
Xiaofei Xie$^1$\and
Yihao Huang$^4$\and
Jian Wang$^1$\and
Yang Liu$^{1,5}$\\ % Felix: should work now :-)
\affiliations
$^1$Nanyang Technological University, Singapore\\
$^2$Alibaba Group, USA\\
$^3$Kyushu University, Japan\\
$^4$East China Normal University, China\\
$^5$Institute of Computing Innovation, Zhejiang University, China\\
}

\begin{document}

\maketitle

\begin{abstract}

In recent years, generative adversarial networks (GANs) and its variants have achieved unprecedented success in image synthesis. They are widely adopted in synthesizing facial images which brings potential security concerns to humans as the fakes spread and fuel the misinformation. However, robust detectors of these AI-synthesized fake faces are still in their infancy and are not ready to fully tackle this emerging challenge. In this work, we propose a novel approach, named \emph{FakeSpotter}, based on monitoring neuron behaviors to spot AI-synthesized fake faces. The studies on neuron coverage and interactions have successfully shown that they can be served as testing criteria for deep learning systems, especially under the settings of being exposed to adversarial attacks. Here, we conjecture that monitoring neuron behavior can also serve as an asset in detecting fake faces since layer-by-layer neuron activation patterns may capture more subtle features that are important for the fake detector. Experimental results on detecting four types of fake faces synthesized with the state-of-the-art GANs and evading four perturbation attacks show the effectiveness and robustness of our approach.
\end{abstract}

%---------------------------------------------------------------------
%---------------------------------------------------------------------
\section{Introduction}

\begin{figure}[t]
	\centering
	\includegraphics[width=1\columnwidth]{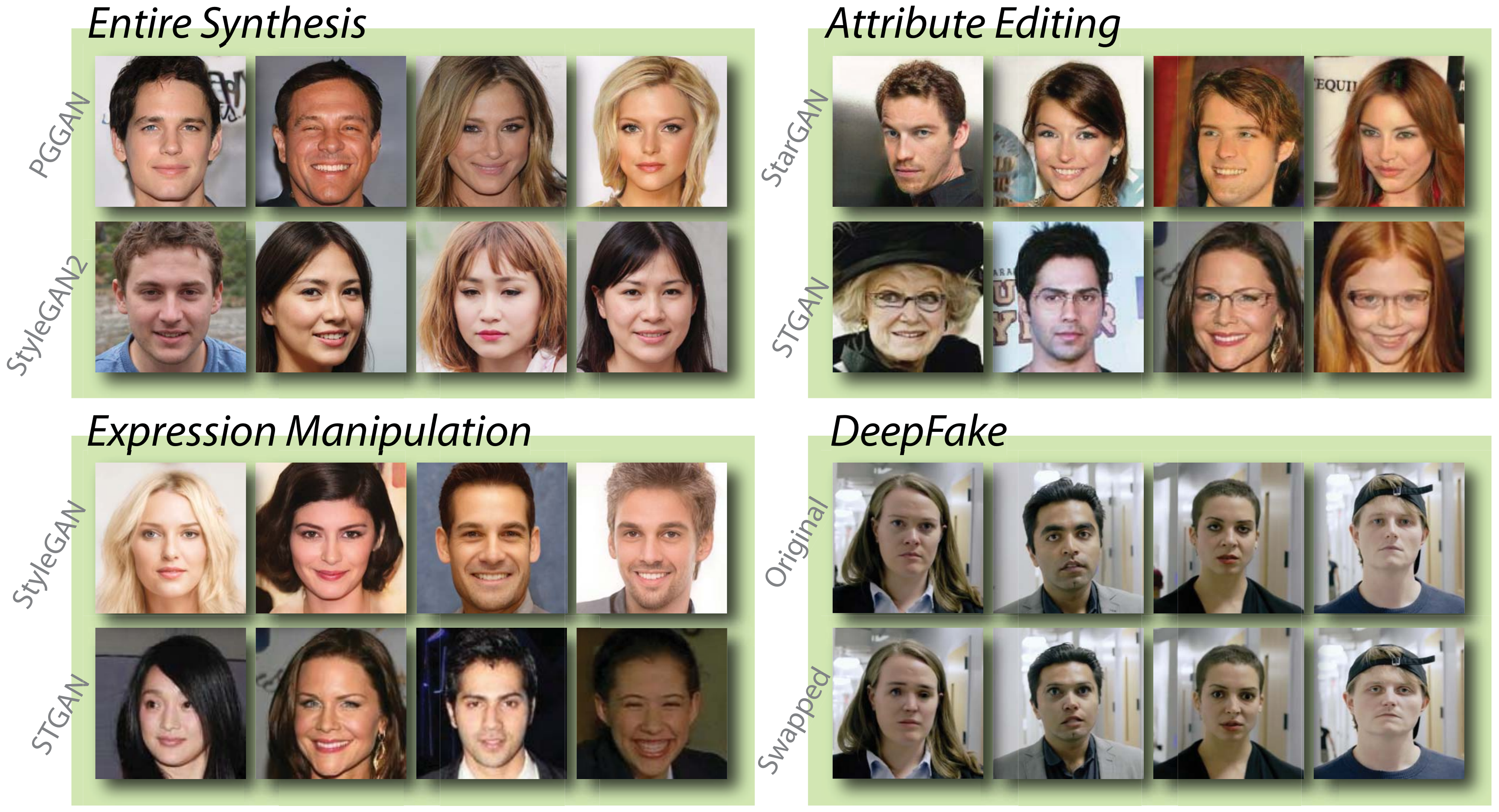} 
	\caption{Four types of fake faces synthesized with various GANs. For the entire synthesis, the facial images are non-existent faces in the world. For attribute editing, StarGAN changes the color of hair into brown and STAGN wears eyeglasses. For expression manipulation, both StyleGAN and STGAN manipulate the face with a smile expression. For DeepFake, the data is from the DeepFake dataset in FaceForensics++ \protect \cite{roessler2019faceforensicspp} and they involve face swap.}
	\label{Figure:fig1}
\end{figure}

With the remarkable development of AI, particularly GANs, seeing is no longer believing nowadays. GANs (\eg, StyleGAN \cite{karras2019style}, STGAN \cite{liu2019stgan}, and StarGAN \cite{choi2018stargan}) exhibit powerful capabilities in synthesizing human imperceptible fake images and editing images in a natural way. Humans can be easily fooled by these synthesized fake images\footnote{\url{https://thispersondoesnotexist.com}}. Figure \ref{Figure:fig1} presents four typical fake faces synthesized with various GANs, which are really hard for humans to distinguish at the first glance. 
%----------------------

The AI-synthesized fake faces not only bring fun to users but also raise security and privacy concerns and even panics to everyone including celebrities, politicians, \etc. Some apps (\eg, FaceApp, Reflect, and ZAO) employ face-synthesis techniques to provide attractive and interesting services such as face swap, facial expression manipulation with several taps on mobile devices. Unfortunately, abusing AI in synthesizing fake images raises security and privacy concerns such as creating fake pornography \cite{porn-pornography}, where a victim's face can be naturally swapped into a naked body and indistinguishable to humans' eyes with several photos \cite{zakharov2019few}. Politicians will also be confused by fake faces, for example, fake official statements may be announced with nearly realistic facial expressions and body movements by adopting AI-synthesized fake face techniques. Due to the potential and severe threats of fake faces, it is urgent to call for effective techniques to spot fake faces in the wild. In this paper, the AI-synthesized fake face or fake face means that the face is synthesized with GANs unless particularly addressed.

\textbf{Entire face synthesis}, \textbf{facial attribute editing}, \textbf{facial expression manipulation}, and \textbf{DeepFake} are the four typical fake face synthesis modes with various GANs \cite{stehouwer2019detection}. Entire face synthesis means that a facial image can be wholly synthesized with GANs and the synthesized faces do not exist in the world. Facial attribute editing manipulates single or several attributes in a face like hair, eyeglass, gender, \etc. Facial expression manipulation alters one's facial expression or transforms facial expressions among persons. DeepFake is also known as the identity swap. It normally swaps synthesized face between different persons and is widely applied in producing fake videos \cite{agarwal2019protecting}. More recently, there is some work that starts to study this topic. However, none of the previous approaches fully tackle the aforementioned four types of fake faces and thoroughly evaluate their robustness against perturbation attack with various transformations in order to show their potentials in dealing with fakes in the wild.

In this paper, we propose a novel approach, named \emph{FakeSpotter}, which detects fake faces by monitoring neuron behaviors of deep face recognition (FR) systems with a simple binary-classifier. Specifically, FakeSpotter leverages the power of deep FR systems in learning the representations of faces and the capabilities of neurons in monitoring the layer-by-layer behaviors which can capture more subtle differences for distinguishing between real and fake faces.

To evaluate the effectiveness of FakeSpotter in detecting fake faces and its robustness against perturbation attacks, we collect numerous high-quality fake faces produced with the state-of-the-art (SOTA) GANs. For example, our entire synthesized fake faces are generated with 1) the freshly released StyleGAN2 \cite{karras2019analyzing}, 2) the newest STGAN \cite{liu2019stgan} that performs facial attributes editing, 3) \textit{DeepFake} that is composed of public datasets (\eg, FaceForensics++ and \textit{Celeb-DF} \cite{Li2019celebdf}), and 4) the Facebook announced real-world DeepFake detection competition dataset (\ie, DFDC). Experiments are evaluated on our collected four types of high-quality fake faces and the results demonstrate the effectiveness of FakeSpotter in spotting fake faces and its robustness in tackling four perturbation attacks (\eg, \textbf{adding noise}, \textbf{blur}, \textbf{compression}, and \textbf{resizing}). FakeSpotter also outperforms prior work AutoGAN \cite{zhang2019detecting} and gives an average detection accuracy of more than 90\% on the four types of fake faces. The average performance measured by the AUC score is merely down less than 3.77\% in tackling the four perturbation attacks under various intensities. 

Our main contributions are summarized as follows.
%-------------------------
\begin{itemize}[leftmargin=*]
    \item \textbf{New observation of neurons in spotting AI-synthesized fake faces.} We observe that layer-by-layer neuron behaviors can be served as an asset for distinguishing fake faces. Additionally, they are also robust against various perturbation attacks at various magnitudes. 
	\item \textbf{Presenting a new insight for spotting AI-synthesized fake faces by monitoring neuron behaviors.} We propose the first neuron coverage based fake detection approach that monitors the layer-by-layer neuron behaviors in deep FR systems. Our approach provides a novel insight for spotting AI aided fakes with neuron coverage techniques.
	\item \textbf{Performing the first comprehensive evaluation on four typical AI-synthesized fake faces and robustness against four common perturbation attacks.} Experiments are conducted on our collected high-quality fake faces synthesized with the SOTA GANs and real dataset like DFDC. Experimental results have demonstrated the effectiveness and robustness of our approach.
\end{itemize}

%---------------------------------------------------------------------
%---------------------------------------------------------------------
\section{Related Work} \label{sec:related work}

%---------------------------------------------------------------------
\subsection{Image Synthesis}
GANs have made impressive progress in image synthesis \cite{zhu2017unpaired,yi2017dualgan} which is the most widely studied area of the applications of GANs since it is first proposed in 2014 \cite{goodfellow2014generative}. The generator in GANs learns to produce synthesized samples that are almost identical to real samples, while the discriminator learns to differentiate between them. Recently, various GANs are proposed for facial image synthesis and manipulation.

In entire face synthesis, PGGAN \cite{karras2017progressive} and StyleGAN, created by NVIDIA, produce faces in high resolution with unprecedented quality and synthesize non-existent faces in the world. STGAN and StarGAN focus on face editing which manipulates the attributes and expressions of humans' faces, \eg, changing the color of hair, wearing eyeglasses, and laughing with a smile or showing feared expression, \etc. \textit{FaceApp} and \textit{FaceSwap} employ GANs to generate \textit{DeepFake} which involves identity swap.

Currently, GANs can be well applied in synthesizing entire fake faces, editing facial attributes, manipulating facial expressions, and swapping identities among persons (\emph{a.k.a.} DeepFake). Fake faces synthesized with the SOTA GANs are almost indistinguishable to humans in many cases. We are living in a world where we cannot believe our eyes 
anymore.

%---------------------------------------------------------------------
\subsection{Fake Face Detection}
Some researchers employ traditional forensics-based techniques to spot fake faces/images. These work inspect the disparities in pixel-level between real and fake images. However, they are either susceptible to perturbation attacks like compression that is common in producing videos with still images \cite{bohme2013counter}, or do not scale well with the increasing amount of training data, as commonly found in shallow learning-based fake detection methods such as \cite{icip16_paint}. Another line in detecting fake images is leveraging the power of deep neural networks (DNNs) in learning the differences between real and fake which are also vulnerable to perturbation attacks like adding human-imperceptible noise \cite{Ian2015ehad}.

In forensics-based fake detection, Nataraj \etal \cite{nataraj2019detecting} employ a DNN model to learn the representation in order to compute co-occurrence matrices on the RGB channels. McCloskey \etal \cite{mccloskey2018detecting} observe that the frequency of saturated pixels in GAN-synthesized fake images is limited as the generator's internal values are normalized and the formation of a color image is vastly different from real images which are sensitive to spectral analysis. Different from forensics-based fake detection, Stehouwer \etal \cite{stehouwer2019detection} introduce an attention-based layer in convolutional neural networks (CNNs) to improve fake identification performance. Wang \etal \cite{wang2019cnn} use ResNet-50 to train a binary-classifier for CNN-synthesized images detection. AutoGAN \cite{zhang2019detecting} trains a classifier to identify the artifacts inducted in the up-sampling component of the GAN.

Other work explores various \textit{ad-hoc} features to investigate artifacts in images for differentiating real and synthesized facial images. For example, mismatched facial landmark points \cite{yang2019exposing}, fixed size of facial area \cite{li2018exposing}, and unique fingerprints of GANs \cite{zhang2019detecting,yu2019attributing}, \etc. These approaches will be invalid in dealing with improved or advanced GANs. Existing works are sensitive to perturbation attacks, but robustness is quite important for a fake detector deployed in the wild.

%---------------------------------------------------------------------
%---------------------------------------------------------------------
\section{Our Method}\label{sec:approach}

In this section, we first give our basic insight and present an overview of FakeSpotter in spotting fake faces by monitoring neuron behaviors. Then, a neuron coverage criteria mean neuron coverage (\textit{MNC}) is proposed for capturing the layer-by-layer neuron activation behaviors. Finally, FakeSpotter differentiates four different types of fake faces with a simple binary-classifier.

%---------------------------------------------------------------------
\subsection{Insight}
% \felix{here}
Neuron coverage techniques are widely adopted for investigating the internal behaviors of DNNs and play an important role in assuring the quality and security of DNNs. It explores activated neurons whose output values are larger than a threshold. The activated neurons serve as another representation of inputs that preserves the learned layer-by-layer representations in DNNs.
Studies have shown that activated neurons exhibit strong capabilities in capturing more subtle features of inputs that are important for studying the intrinsic of inputs. DeepXplore \cite{pei2017deepxplore} first introduces neuron coverage as metrics for DNN testing to assure their qualities. Some work exploits the critical activated neurons in layers to detect adversarial examples for securing DNNs \cite{ma2019nic,ma2019deepct,ma2018deepmutation,zhang2019machine}.

\begin{figure}[t]
	\centering
	\includegraphics[width=1\columnwidth]{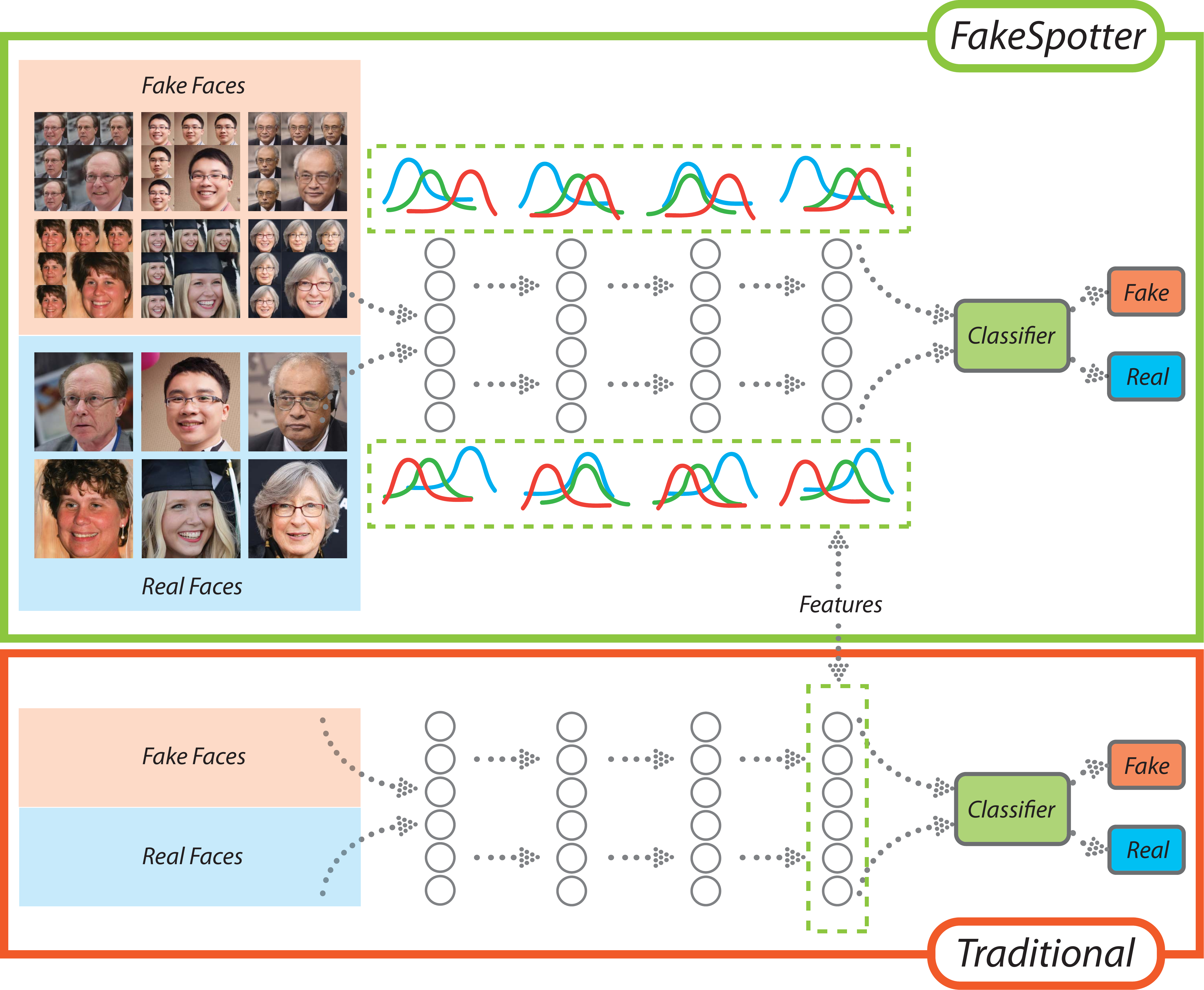} 
	\caption{An overview of the proposed fake face detection method, FakeSpotter. Compared to the traditional learning-based method (shown at the bottom), the FakeSpotter uses layer-wise neuron behavior as features, as opposed to final-layer neuron output. Our approach uses a shallow neural network as the classifier while traditional methods rely on deep neural networks in classification.}
	\label{Figure:fig2}
\end{figure}

Our work is motivated by the power of layer-wise activated neurons in capturing the subtle features of inputs which could be used for amplifying the differences between real and synthesized facial images. Based on this insight, we propose FakeSpotter by monitoring the neuron behaviors in deep FR systems (\eg, VGG-Face) for fake face detection. Deep FR systems have made incredible progress in face recognition but are still vulnerable to identifying fake faces \cite{korshunov2018deepfakes}. In Figure \ref{Figure:fig2}, we present an overview of FakeSpotter using layer-wise neuron behavior as features with a simple binary-classifier to identify real and fake faces.
%---------------------------------------------------------------------

%---------------------------------------------------------------------
%%\vspace{-10pt}
\subsection{Monitoring Neuron Behaviors}
In DNNs, a neuron is a basic unit and the final layer neuron outputs are employed for prediction. Given an input of trained DNN, the activation function $\phi$ (\eg, Sigmoid, ReLU) computes the output value of neurons with connected neurons $\mathrm{x}_i$ in the previous layers, weights matrix $W^k_i$, and bias $b_j$. Activated neurons in each individual layers are determined by whether the output value is higher than a threshold $\xi$.

In this work, we propose a new neuron coverage criterion, named mean neuron coverage (\textit{MNC}), for determining the threshold $\xi$. Existing approaches \cite{ma2018deepgauge,xie2019deephunter} in calculating threshold $\xi$ are mostly designed for testing DNNs and are not applicable for fake detection. Pei \etal \cite{pei2017deepxplore} define a global threshold for activating neurons in all layers, which is too rough.

In DNNs, each layer plays their own unique roles in learning the representations of inputs \cite{Mahendran_2015_CVPR}. Here, we introduce another strategy by specifying a threshold $\xi_l$ for each layer $l$. The threshold $\xi_l$ is the average value of neuron outputs in each layer for given training inputs. The layer $l$ is the convolutional and fully-connected layers which are valuable layers preserving more representation information in the model. Specifically, we calculate the threshold $\xi_l$ for each layer with the following formula:
\begin{equation} \label{eq:1}
\xi_l=\frac{\sum_{n \in N, t \in \mathcal{T}}^{}\delta(n,t)}{|N| \cdot |\mathcal{T}|}
\end{equation}
where $N$ represents a set of neurons in the $l$th layer and $|N|$ is the total number of neurons in the $N$, $\mathcal{T}$=$\{t_1,t_2,...,t_k\}$ is a set of training inputs and $|\mathcal{T}|$ indicates the number of training inputs in $\mathcal{T}$, $\delta(n,t)$ calculates the neurons output value where $n$ is the neuron in $N$ and $t$ denotes the input in $\mathcal{T}$. Finally, our neuron coverage criterion \textit{MNC} determines whether a neuron in the $l$th layer is activated or not by checking whether its output value is higher than the threshold $\xi_l$. We define the neuron coverage criterion \textit{MNC} for each layer $l$ as follows:
\begin{equation} \label{eq:2}
\mathrm{\emph{MNC}}(l, t)=|\{{n|\forall n\in l, \delta(n, t)>\xi_l}\}|
\end{equation}
where $t$ represents the input, $n$ is the neuron in layer $l$, $\delta$ is a function for computing the neuron output value, and $\xi_l$ is the threshold of the $l$-th layer calculated by formula (\ref{eq:1}).

%---------------------------------------------------------------------
\subsection{Detecting Fake Faces}

As described above, we capture the layer-wise activated neurons with \textit{MNC}. We train a simple binary-classifier with shallow neural networks. The input of our classifier is the \textit{general} neuron behavior rather than the \textit{ad-hoc} raw pixels like traditional image classification models. Raw pixels could be easily perturbed by attackers and trigger erroneous behaviors.

Algorithm~\ref{alg:procedure} describes the procedure of fake face detection. First, the thresholds for determining neuron activation in each layer are identified by our proposed neuron coverage criterion \textit{MNC} with fake and real faces as the training dataset, denoted as $\mathcal{T}$. Then, a feature vector for each input face is formed as the number of activated neurons in each layer. Let $\mathit{F}$=$\{f_1,f_2,...,f_i,...,f_m\}$ and $\mathit{R}$=$\{r_1,r_2,...,r_j,...,r_m\}$ represent the feature vector of fake and real input faces respectively, where $f_i$ and $r_j$ are the number of activated neurons in the $i$th and $j$th layer, $m$ is the total number of layers in deep FR system. Finally, we train a supervised binary-classifier, denoted as $\widetilde{C}$, by receiving the formed feature vectors of fake and real faces as inputs to predict the input is real or fake. 
% \fei{How to determine the thresholds?}
% \setlength{\textfloatsep}{0pt}% Remove \textfloatsep
\begin{algorithm}[t]
% %\vspace{-10pt}
% 	\footnotesize
    % \scriptsize
	\SetAlgoLined
	\SetKwInOut{Input}{Input}
	\SetKwInOut{Output}{Output}
	\Input{Training dataset of fake and real faces $\mathcal{T}$, Test dataset of fake and real faces $\mathcal{D}$, Pre-trained deep FR model $\widetilde{M}$}
	\Output{Label $tag$}

	$\mathrm{L}$ is the convolutional and fully-connected layers in $\widetilde{M}$.\\
	$\triangleright$ Determine the threshold of neuron activation for each layer.\\
	
	\For{$t\in \mathcal{T}$}{
	$\mathrm{N}$ is a set of neurons in the $l$th layer of $\widetilde{M}$.\\
	$\mathrm{S}$ saves neuron output value for a given input $t$.\\
   			\For {$l \in \mathrm{L}$, $n \in \mathit{N}$}{
					$\mathrm{S}_l = \sum_{} \delta(n,t)$
		   }
	      $\xi_l=\frac{1}{|L|}\cdot{\mathrm{S}}$
	}
   $\triangleright$ Train a binary-classifier for detecting fake/real faces.\\
   $\mathrm{V}$ counts activated neurons in $\mathrm{L}$.\\
   \For{$t\in \mathcal{T}$}{
   			\For {$l \in \mathrm{L}$, $n \in \mathit{N}$}{
   			    \If{$\delta(n,t)>\xi_{l}$}{
   					$\mathrm{V}_l \leftarrow n$
   				}
   			}
   }
    Train a binary-classifier $\widetilde{C}$ with inputs $\mathrm{V}$.\\
  
    $\triangleright$ Predict whether a face from test dataset $\mathcal{D}$ is real or fake.\\
  \For{$d\in \mathcal{D}$}{
  		$tag \leftarrow \mathrm{argmax} \ \widetilde{C}(d)$\\
}
		\Return $tag$
		\caption{Algorithm for detecting fake faces with neuron coverage in deep FR systems.}
		\label{alg:procedure}
\end{algorithm}
% %\vspace{-5pt}
% \setlength{\textfloatsep}{15pt}

In prediction, an input face should be processed by a deep FR system to extract the neuron coverage behaviors with our proposed criterion \textit{MNC}, namely the number of activated neurons in each layer. The activated neurons are formed as a feature to represent the input face. Then, the trained binary-classifier predicts whether the input is a real or fake face.

%---------------------------------------------------------------------
%---------------------------------------------------------------------
\section{Experiments} \label{sec:eval}

In this section, we conduct experiments to evaluate the effectiveness of FakeSpotter in spotting four types of fake faces produced with the SOTA techniques and investigate its robustness against four common perturbation attacks. We present the experimental results of detection performance with a comparison of recently published work AutoGAN \cite{zhang2019detecting} in Section \ref{Sec:dp} and robustness analysis in Section \ref{Sec:ra}. In Section \ref{section:cele_DF}, we provide the comparison results in detecting a public DeepFake dataset \textit{Celeb-DF}.

%---------------------------------------------------------------------
\subsection{Experimental Setup}
% $\bullet$ \textbf{Data Collection.} 
\paragraph{Data Collection.} In our experiments, real face samples are collected from CelebA \cite{liu2015faceattributes} and Flicker-Faces-HQ (FFHQ) since they exhibit good diversity. We also utilize original real images provided by the public dataset FaceForensics++, DFDC\footnote{\footnotesize DeepFakes Detection Challenge (DFDC) Dataset by Facebook. \url{https://www.kaggle.com/c/DeepFake-detection-challenge}}, and \textit{Celeb-DF}.
To ensure the diversity and high-quality of our fake face dataset, we use the newest GANs for synthesizing fake faces (\eg, StyleGAN2) using the public dataset (\eg, \textit{Celeb-DF}), and real dataset such as DFDC dataset. The DFDC dataset is the officially released version rather than the preview edition. Table \ref{Table:data_collection} presents the statistics of our collected fake face dataset.
% -----------------------------
\begin{table}[t]
\scriptsize
\centering
\setlength{\tabcolsep}{3.5pt}
\begin{tabular}{c|c|c|c|c}
\toprule
Fake Faces                         & GAN Type               & Manipulation & Real Source & Collection   \\ \midrule
\multirow{2}{*}{\makecell[c]{Entire \\ Synthesis}}  & PGGAN     &  full         & CelebA       &  self-synthesis    \\ \cline{2-5} 
                                   & StyleGAN2         &   full       &    FFHQ    & officially released  \\ \midrule
\multirow{2}{*}{\makecell[c]{Attribute \\ Editing} }  & StarGAN    &   brown-hair        &CelebA       & self-synthesis\\ \cline{2-5} 
                                                & STGAN             &    eyeglasses       &  CelebA       & self-synthesis  \\ \midrule
\multirow{2}{*}{\makecell[c]{Expression \\ Manipulation}} & StyleGAN  &    ctrl. smile intensity  &  FFHQ  &  self-synthesis \\ \cline{2-5} 
                                   & STGAN           &   smile       &  CelebA      & self-syntheis     \\ \midrule
\multirow{3}{*}{\makecell[c]{DeepFake}}  & F. F. ++   &   face swap        &    unknown    & FaceForensics++   \\ \cline{2-5} 
                                   & DFDC       &  face/voice swap      &   unknown     & Kaggle dataset   \\ \cline{2-5}
                                   &  Celeb-DF  & face swap  & YouTube  & Celeb-DF(V2) \\
\bottomrule
\end{tabular}
\caption{Statistics of collected fake faces dataset. Column \textit{Manipulation} indicates the manipulated region in face. Column \textit{Real Source} denotes the source of real face for producing fake faces. Last column \textit{Collection} means the way of producing fake faces, synthesized by ourselves or collected from public dataset. \textit{F.F. ++} denotes FaceForensics++ dataset.}
\label{Table:data_collection}
%\vspace{-5pt}
\end{table}

% \noindent$\bullet$ \textbf{Implementation Details.}
\paragraph{Implementation Details.}We design a shallow neural network with merely five fully-connected layers as our binary-classifier for spotting fakes. The optimizer is SGD with momentum 0.9 and the starting learning rate is 0.0001, with a decay of $1e$-$6$. The loss function is binary cross-entropy.

In monitoring neuron behaviors with \textit{MNC}, we utilize VGG-Face\footnote{\url{https://github.com/rcmalli/keras-vggface}} with ResNet50 as backend architecture for capturing activated neurons as it can well balance detection performance and computing overhead. Our approach is generic to FR systems, which could be easily extended to other deep FR systems. In evaluating the robustness in tackling perturbation attacks, we select four common transformations, namely \textit{compression}, \textit{resizing}, \textit{adding noise}, and \textit{blur}.

% \noindent$\bullet$ \textbf{Training and Test Dataset.}
\paragraph{Training and Test Dataset.} Using the training dataset $\mathcal{T}$, we train the model with 5,000 real and 5,000 fake fakes for each individual GAN. In the test dataset $\mathcal{D}$, we use 1,000 real and 1,000 fake faces for evaluation. The training and test dataset are based on different identities. The training dataset $\mathcal{T}$ and test dataset $\mathcal{D}$ are employed for evaluating the effectiveness and robustness of FakeSpotter. The \textit{Celeb-DF} dataset provides another independent training and test dataset for comparing the performance with existing thirteen methods in detecting fake videos on \textit{Celeb-DF}.

% \noindent$\bullet$ \textbf{Evaluation Metrics.}
\paragraph{Evaluation Metrics.} In spotting real and fake faces, we adopt eight popular metrics to get a comprehensive performance evaluation of FakeSpotter. Specifically, we report precision, recall, F1-score, accuracy, AP (average precision), AUC (area under curve) of ROC (receiver operating characteristics), FPR (false positive rate), and FNR (false negative rate), respectively. We also use the AUC as a metric to evaluate the performance of FakeSpotter in tackling various perturbation attacks.

All our experiments are conducted on a server running Ubuntu 16.04 system on a total 24 cores 2.20GHz Xeon CPU with 260GB RAM and two NVIDIA Tesla P40 GPUs with 24GB memory for each.

%---------------------------------------------------------------------
\subsection{Detection Performance} \label{Sec:dp}

In evaluating the performance of FakeSpotter in detecting fake faces and its generalization to different GANs. We select four totally different types of fake faces synthesized with various GANs and compare with prior work AutoGAN. To get a comprehensive performance evaluation, we use eight different metrics to report the detection rate and false alarm rate. 

\begin{figure}
\centering
\includegraphics[width=\columnwidth]{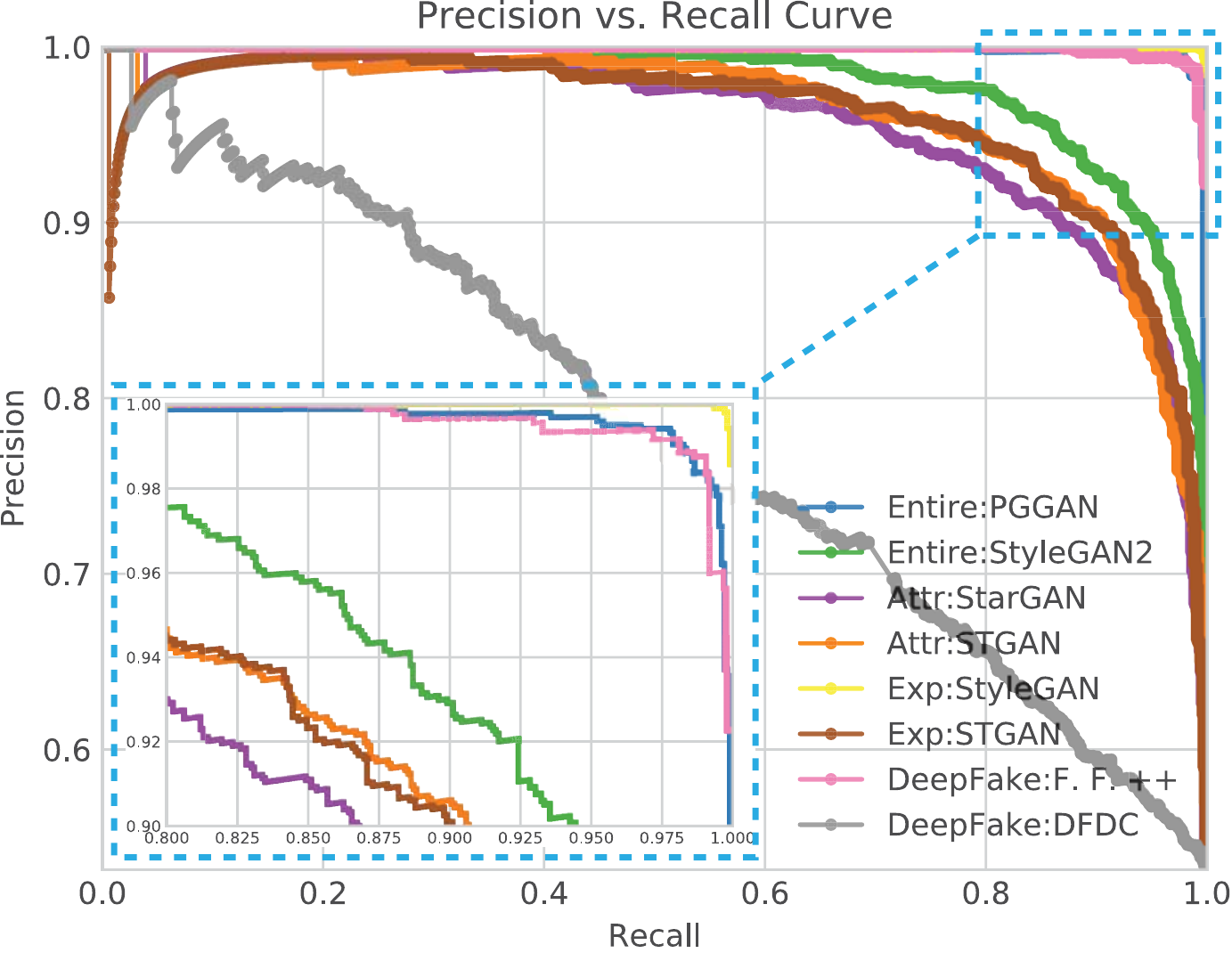}
\caption{Precision-recall curves of the four types of fake faces. The curve computes precision-recall for different probability thresholds.}
\label{Figure:PR_performance}
\end{figure}

Table \ref{Table:performance} shows the performance of FakeSpotter and prior work AutoGAN in detecting fake faces measured by eight different metrics. AutoGAN is a recent open source work that leverages the artifacts existed in GAN-synthesized images and detects the fake image with a deep neural network-based classifier. Furthermore, to illustrate the performance of FakeSpotter in balancing the precision and recall, we present the precision and recall curves in Figure \ref{Figure:PR_performance} as well.
% -----------------------------
\begin{table*}[t]
\scriptsize
\centering
% \caption{Performance of FakeSpotter (\emph{F. S.}) and AutoGAN (\emph{A. G.}) in spotting the four types of fake faces. PGGAN and StyleGAN2 produce entire synthesized facial images. In attribute editing, StarGAN manipulates the color of the hair with brown, STGAN manipulates face by wearing eyeglasses. In Expression manipulation, StyleGAN and STGAN manipulate the expression of faces with the smile while StyleGAN can control the intensity of the smile. Average performance is an average results over the fake faces. Here, we provide two kinds of average performance, average performance on still images (including the first three types of fake faces) and all the four types of fake faces.}
\setlength{\tabcolsep}{3.5pt} % Felix
\begin{tabular}{c|c|c|c|c|c|c|c|c|c|c|c|c|c|c|c|c|c}
\toprule
 \multirow{2}{*}{Fake Faces} & \multirow{2}{*}{GAN}               & \multicolumn{2}{c|}{precision} & \multicolumn{2}{c|}{recall} & \multicolumn{2}{c|}{F1} & \multicolumn{2}{c|}{accuracy} & \multicolumn{2}{c|}{AP} & \multicolumn{2}{c|}{AUC} & \multicolumn{2}{c|}{FPR} & \multicolumn{2}{c}{FNR}  \\ \cline{3-18}
                     &                & F. S. & A. G. & F. S. & A. G.& F. S. & A. G. & F. S. & A. G. & F. S. & A. G. & F. S. & A. G. & F. S. & A. G. & F. S. & A. G.  \\ \midrule
\multirow{2}{*}{Entire Synthesis}  & PGGAN             & 0.986  & 0.926  & 0.987 & 0.974 & 0.986 & 0.949 & 0.986 & 0.948 & 0.979 & 0.915 & 0.985 & 0.948 & 0.013 & 0.026  & 0.016  &  0.078     \\ \cline{2-18} 
                                   & StyleGAN2         & 0.912  & 0.757  & 0.924 & 0.663 & 0.918 & 0.707 & 0.919 & 0.725 & 0.881 & 0.670 & 0.919 & 0.725 & 0.076 & 0.337  & 0.087  &  0.213     \\ \midrule
\multirow{2}{*}{Attribute Editing} & StarGAN           &  0.901  & 0.690  &  0.865 & 0.567 &  0.883 & 0.622  & 0.88 & 0.656 & 0.851 & 0.608  & 0.881 & 0.656 & 0.135 & 0.433  & 0.104 &  0.255      \\ \cline{2-18} 
                                   & STGAN             & 0.885  & 0.555  & 0.918 & 0.890 & 0.901 & 0.683 & 0.902 & 0.588 &  0.852 & 0.549 & 0.902 & 0.588 & 0.082 & 0.11  & 0.114  &  0.715    \\ \midrule
\multirow{2}{*}{\makecell[c]{Expression \\ Manipulation}} & StyleGAN  &  1.0   & 0.736     & 0.983 & 0.920 & 0.991 & 0.818   & 0.991 & 0.795 & 0.992 & 0.717   & 0.991 & 0.795 & 0.017 & 0.08  & 0.0  &  0.33      \\ \cline{2-18} 
                                   & STGAN             &  0.898  & 0.0 & 0.913 & 0.0 & 0.905 & 0.0 &  0.906 & 0.5 & 0.863  & 0.5  &   0.906 & 0.5  & 0.087 & 1.0  & 0.102    & 0.0    \\ \midrule
\multirow{2}{*}{DeepFake}          & FaceForensics++   &  0.978  & 0.508  & 0.992 & 0.629 & 0.985 & 0.562 &  0.985 & 0.511 & 0.973 & 0.505 & 0.985 & 0.511  &  0.008 & 0.371 &  0.021 & 0.608      \\ \cline{2-18} 
                                   & DFDC              &  0.691 &  0.536 & 0.719 & 1.0 & 0.705 & 0.698 & 0.682 & 0.536 & 0.645  & 0.536 & 0.680 & 0.5  & 0.281 & 0.0   & 0.359 & 1.0      \\ \midrule
\multicolumn{2}{c|}{Average Performance (first three types)} & 0.930   & 0.611  & 0.932 & 0.669 & 0.931 & 0.630  & 0.931   & 0.702  & 0.903  & 0.660 & 0.931  & 0.702  & 0.068  & 0.331  & 0.071  & 0.265 \\ \midrule
\multicolumn{2}{c|}{Average Performance (all four types)} & 0.906   & 0.589  & 0.913 & 0.705 & 0.909 & 0.630  & 0.906   & 0.657  & 0.880  & 0.625 & 0.906  & 0.653  & 0.087  & 0.295  & 0.10  & 0.40 \\
\bottomrule
\end{tabular}
\caption{Performance of FakeSpotter (\emph{F. S.}) and AutoGAN (\emph{A. G.}) in spotting the four types of fake faces. PGGAN and StyleGAN2 produce entire synthesized facial images. In attribute editing, StarGAN manipulates the color of the hair with brown, STGAN manipulates face by wearing eyeglasses. In Expression manipulation, StyleGAN and STGAN manipulate the expression of faces with the smile while StyleGAN can control the intensity of the smile. Average performance is an average results over the fake faces. Here, we provide two kinds of average performance, average performance on still images (including the first three types of fake faces) and all the four types of fake faces.}
\label{Table:performance}
%\vspace{-4pt}
\end{table*}
% -----------------------------

Experimental results demonstrate that FakeSpotter outperforms AutoGAN and achieves competitive performance with a high detection rate and low false alarm rate in spotting the four typical fake faces synthesized by GANs. We also find that FakeSpotter achieves a better balance between precision and recall on four types of fake faces from Figure \ref{Figure:PR_performance}. Further, we observe some interesting findings from Table \ref{Table:performance}.

First, fake faces synthesized with advanced GANs are difficult to be spotted by FakeSpotter. For example, in entire synthesis, FakeSpotter detects PGGAN with an accuracy of $98.6\%$, but gives an accuracy of $91.8\%$ on StyleGAN2 (the best performed GAN in entire synthesis and just released by NVIDIA). In addition, entire face synthesis is easily spotted than partial manipulation of fake faces that may contain less fake footprints. These two findings indicate that well-designed GANs and minor manipulations could produce more realistic and harder-to-spot fake faces.

In Table \ref{Table:performance}, the performance of FakeSpotter in detecting DFDC is not ideal as other types of fake faces since fake faces in DFDC could be either a face swap or voice swap (or both) claimed by Facebook. In our experiments, some false alarms could be caused by the voice swap which is out the scope of FakeSpotter. A potential idea of detecting fakes with random face and voice swap combination is inferring the characteristic physical features of faces from voice, and vice versa. 

%---------------------------------------------------------------------
%\vspace{-4pt}
\subsection{Robustness Analysis} \label{Sec:ra}
\begin{figure}[tbp]
\centering
% %\vspace{-0.2cm}  
% \setlength{\abovecaptionskip}{0.2cm}   
% \setlength{\belowcaptionskip}{-0.3cm}   
\subfigure[Compression]{
\includegraphics[width=0.45\columnwidth]{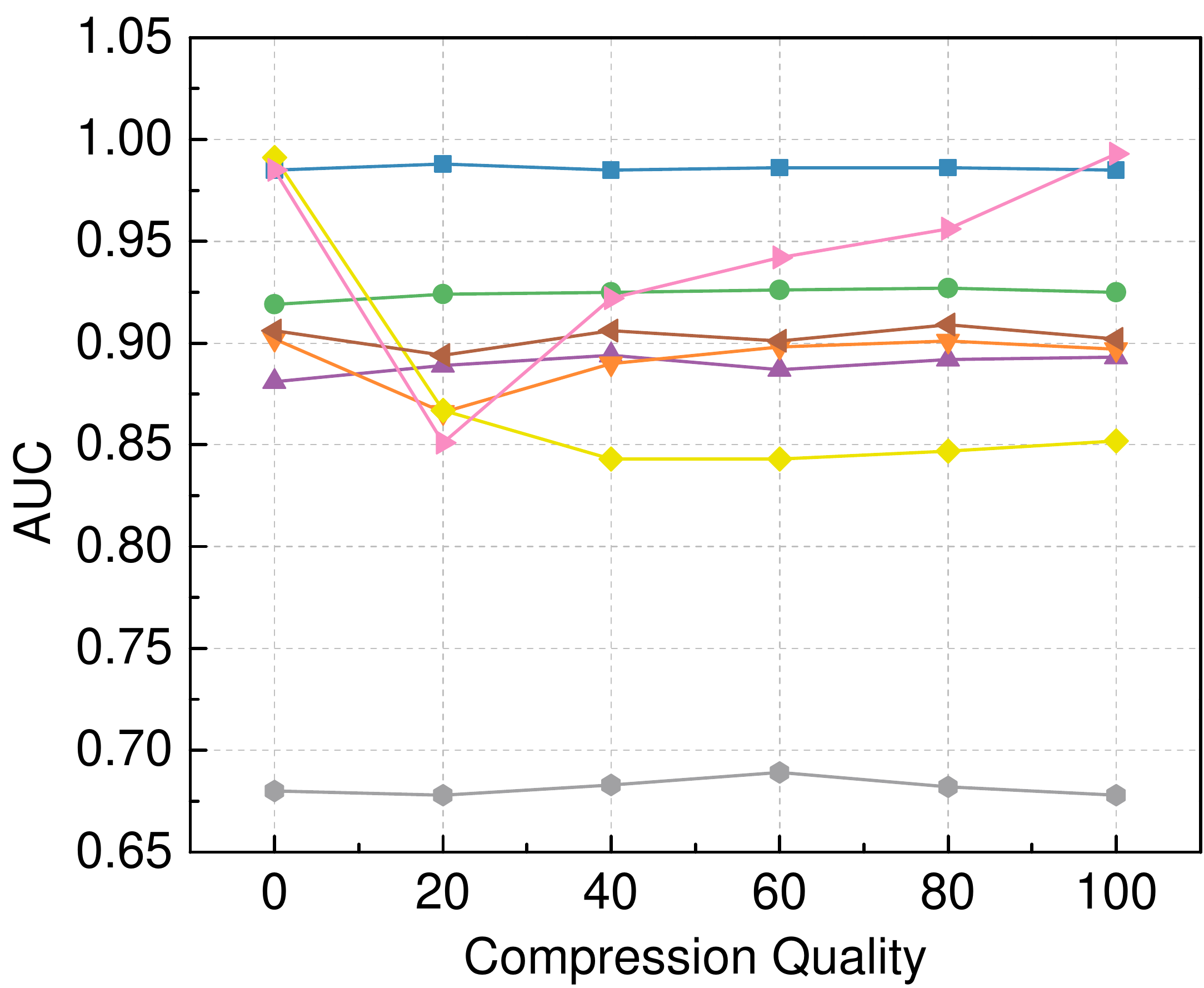}
}
\quad
\subfigure[Blur]{
\includegraphics[width=0.45\columnwidth]{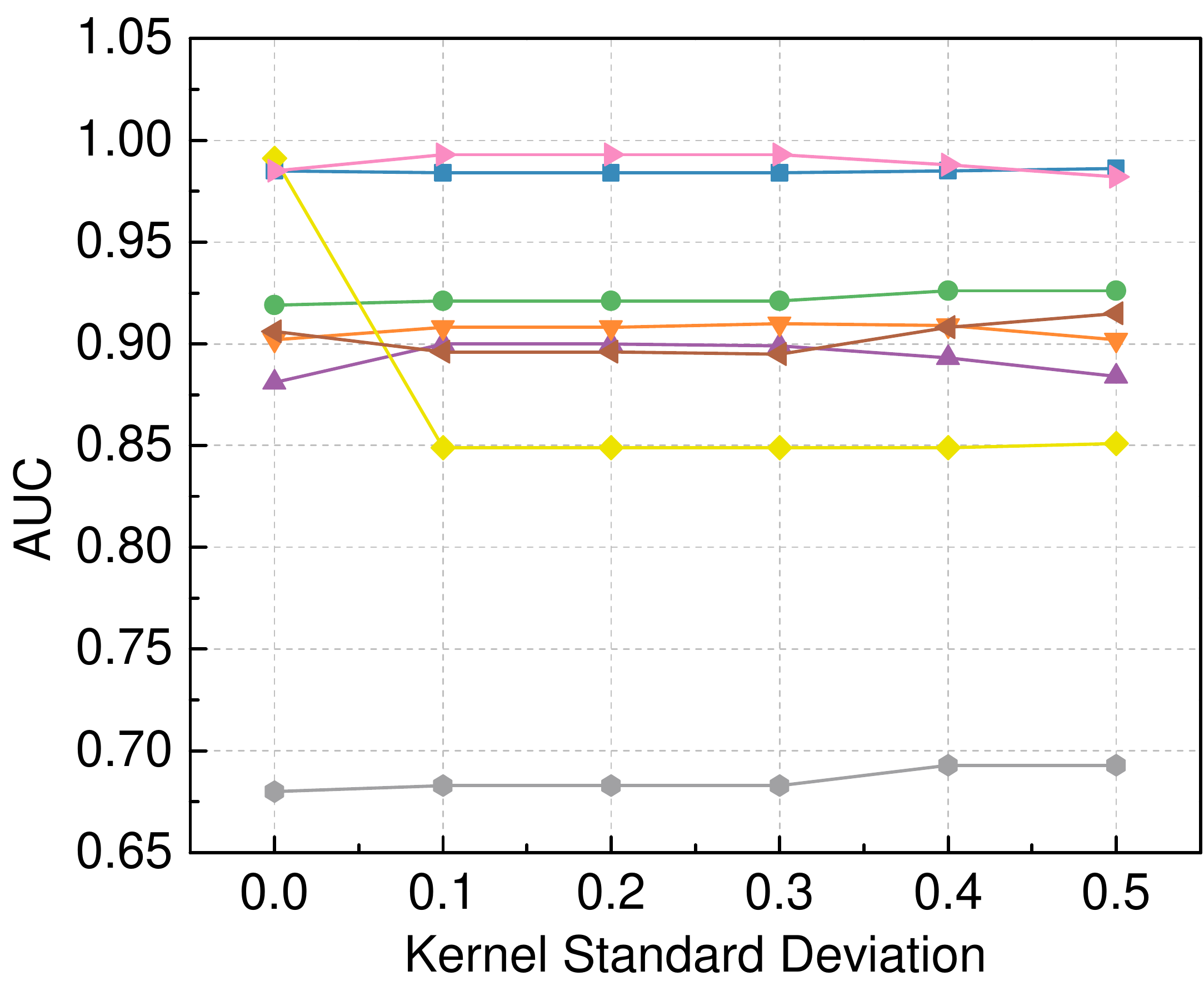}
}
\quad
\subfigure[Resizing]{
\includegraphics[width=0.45\columnwidth]{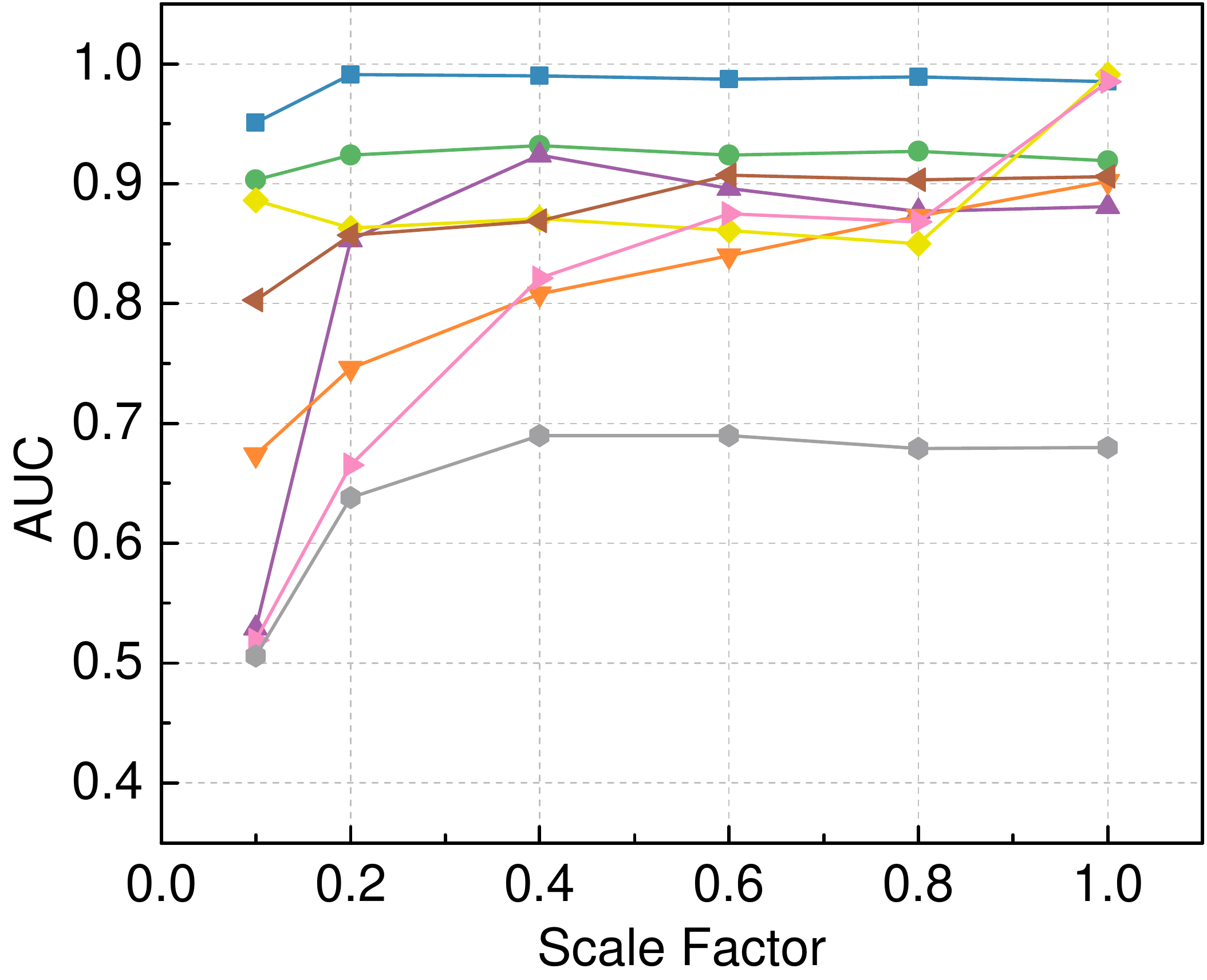}

}
\quad
\subfigure[Noise]{
\includegraphics[width=0.45\columnwidth]{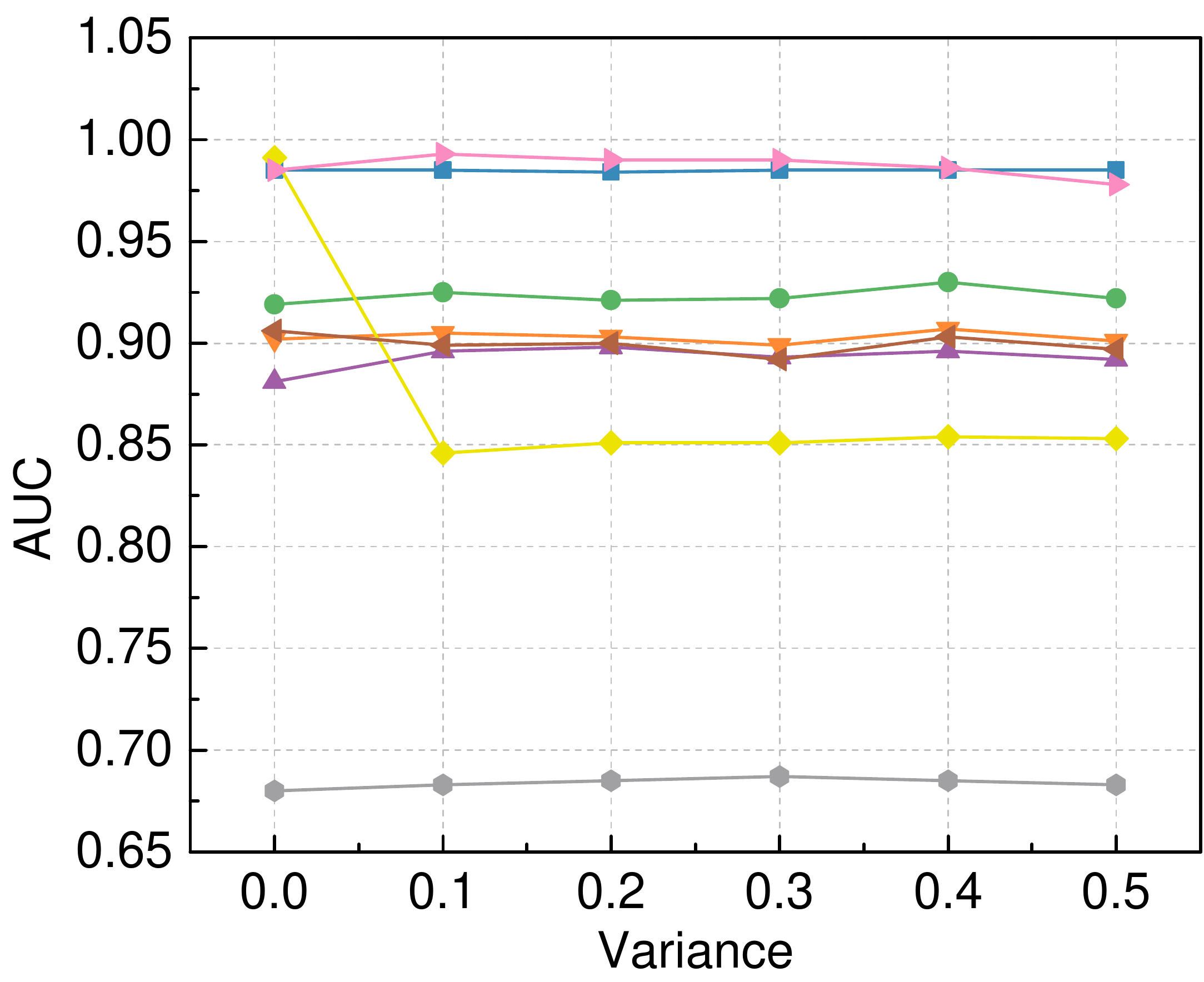}
}
\caption{Four perturbation attacks under different intensities. Legends refer to Figure\ref{Figure:PR_performance}.}
\label{Figure:robusty}
\end{figure}

Robustness analysis aims at evaluating the capabilities of FakeSpotter against perturbation attacks since image transformations are common in the wild, especially in creating fake videos. The transformations should be less sensitive to human eyes. Here, we mainly discuss the performance of FakeSpotter in tackling four different perturbation attacks under various intensities. We utilize the AUC as metrics for the performance evaluation. Figure \ref{Figure:robusty} plots the experimental results of FakeSpotter against the four perturbation attacks.
% -----------------------------

In Figure \ref{Figure:robusty}, the compression quality measures the intensity of compression, the maximum and minimum value are $100$ and $0$, respectively. Blur means that we employ Gaussian blur to faces. The value of Gaussian kernel standard deviation is adjusted to control the intensity of blur while maintaining the Gaussian kernel size to (3, 3) unchanged. In resizing, scale factor is used for controlling the size of an image in horizontal and vertical axis. We add Gaussian additive noise to produce images with noise and the variance is used for controlling the intensity of the noise.

Experimental results demonstrated the robustness of the FakeSpotter in tackling the four common perturbation attacks. We find that the AUC score of FakeSpotter maintains a minor fluctuation range when the intensity of perturbation attacks increased. Due to the severe artifacts in F.F.++ and high intensity of facial expression manipulation in StyleGAN, their variation is a little obvious. The average AUC score of all the four types of fake faces decreased less than 3.77\% on the four perturbation attacks under five different intensities. 

%---------------------------------------------------------------------
%\vspace{-4pt}
\subsection{Performance on \textit{Celeb-DF(v2)}} \label{section:cele_DF}

\textit{Celeb-DF} \cite{Li2019celebdf} is another large-scale DeepFake video dataset with many different subjects (\eg, ages, ethic groups, gender) and contains more than 5,639 high-quality fake videos. In their project website, they provide some comparison results of existing video detection methods on several DeepFake videos including \textit{Celeb-DF}. There are two versions of \textit{Celeb-DF} dataset, \textit{Celeb-DF(v1)} and \textit{Celeb-DF(v2)} dataset, a superset of \textit{Celeb-DF(v1)}.

\begin{figure}[t]
	\centering
	\includegraphics[width=\columnwidth]{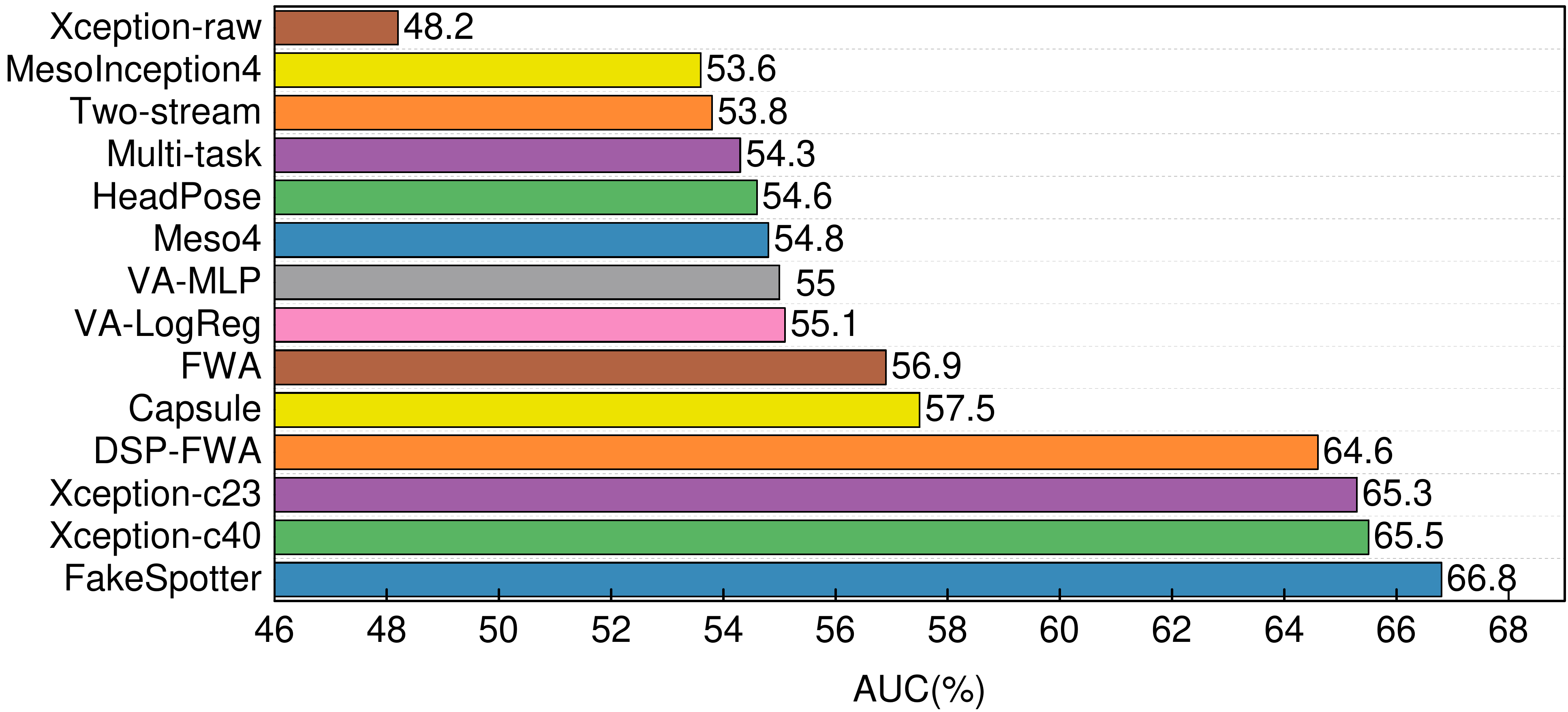}
	\caption{AUC score of various methods on \textit{Cele-DF(V2)} dataset.}
	\label{Figure:fig_celeb_DF}
\end{figure}

We use \textit{Celeb-DF(v2)} dataset for demonstrating the effectiveness of FakeSpotter further and get a more comprehensive comparison with existing work on fake video detection. We also utilize AUC score as metrics for evaluating our approach FakeSpotter as AUC score is served as the metrics in \textit{Celeb-DF} project for comparing with various methods. Figure \ref{Figure:fig_celeb_DF} shows the performance of FakeSpotter in spotting fake videos on \textit{Celeb-DF(v2)}. Experimental results show that FakeSpotter reaches an AUC score 66.8\% on the test dataset provided in \textit{Celeb-DF(v2)} and outperforms all the existing work listed. 

According to the experimental results in Figure \ref{Figure:fig_celeb_DF}, fake video detection is still a challenge, especially when some high-quality fake videos utilize various unknown techniques.
%\vspace{-15pt}
% -----------------------------
% -----------------------------

%---------------------------------------------------------------------
\subsection{Discussion} \label{Sec:diss}

Our approach achieves impressive results in detecting various types of fake faces and is robust against several common perturbation attacks. However, there are also some limitations. The performance of FakeSpotter in spotting DFDC is not as ideal as other types of fake faces. One of the main reasons is that fake faces in DFDC involve two different domain fake, face swap and voice swap. However, our approach only focuses on facial images without any consideration of the voice. This suggests that producing fake multimedia by incorporating various seen and unseen techniques may be a trend in the future. It poses a big challenge to the community and calls for effective approaches to detect these perpetrating fakes. In an adversarial environment, attackers could add adversarial noise to evade our detection, and there is a trade-off between generating imperceptible facial images and the success of evasion.
%\vspace{-8pt}

%---------------------------------------------------------------------
%---------------------------------------------------------------------
\section{Conclusion and Future Research Directions} \label{sec:conclusion}

We proposed the FakeSpotter, the first neuron coverage based approach for fake face detection, and performed an extensive evaluation of the FakeSpotter on fake detection challenges with four typical SOTA fake faces.
FakeSpotter demonstrates its effectiveness in achieving high detection rates and low false alarm rates. In addition, our approach also exhibits robustness against four common perturbation attacks. The neuron coverage based approach presents a new insight for detecting fakes, which we believe could also be extended to other fields like fake speech detection.

Everyone could potentially fall victim to the rapid development of AI techniques that produce fake artifacts (\eg, fake speech, fake videos). The arms race between producing and fighting fakes is on an endless road. Powerful defense mechanisms should be developed for protecting us against AI risks. However, a public database with benchmark containing diverse high-quality fake faces produced by the SOTA GANs is still lacking in the community which could be our future work. In addition, an interplay between our proposed method and novel fake localization methods \cite{arxiv20_fakelocator} is also worth pursuing. Beyond DeepFake detection, we conjecture that the FakeSpotter can work well in tandem with non-additive noise adversarial attacks \eg,  \cite{arxiv19_amora,arxiv20_abba} where the attacked images do not reveal the noise pattern and are much harder to accurately detect. 

% Beyond detection, provenance is another issue that should be considered in fighting fakes. For provenance, we need to answer the following three questions, namely who produced the fake, how to produce the fake, and where the tampered region is in fake. On the attack side, new metrics are needed for evaluating the quality of GANs in image synthesis.

%---------------------------------------------------------------------
%---------------------------------------------------------------------
%\vspace{-8pt}
\section*{Acknowledgments}
% \noindent\textbf{Acknowledgments.}
This research was supported in part by Singapore National Cybersecurity R\&D Program No. NRF2018NCR-NCR005-0001, National Satellite of Excellence in Trustworthy Software System No. NRF2018NCR-NSOE003-0001, NRF Investigatorship No. NRFI06-2020-0022. It was also supported by JSPS KAKENHI Grant No. 20H04168, 19K24348, 19H04086, and JST-Mirai Program Grant No. JPMJMI18BB, Japan. We gratefully acknowledge the support of NVIDIA AI Tech Center (NVAITC) to our research.

% \clearpage
\footnotesize
\bibliographystyle{named}
% \bibliography{ref}
\bibliography{ref_acronym} % Felix: shrink for extra space, now we can add some more references.

% \balance

\end{document}